\documentclass{IEEEcsmag}

\usepackage[colorlinks,urlcolor=blue,linkcolor=blue,citecolor=blue]{hyperref}

\usepackage{upmath}

\jvol{XX}
\jnum{XX}
\paper{8}

\begin{document}


\title{Serverless Predictions: 2021-2030}

\author{Pedro Garcia Lopez}
\affil{Universitat Rovira i Virgili/IBM Watson Research, Spain/USA
}

\author{Aleksander Slominski}
\affil{IBM Watson Research, USA
}

\author{Michael Behrendt}
\affil{IBM Deutschland Research \& Development GmbH, DE}

\author{Bernard Metzler}
\affil{IBM Research Zurich, CH}





\begin{abstract}

Within the next 10 years, advances on resource disaggregation will enable full transparency for most Cloud applications: to 
run unmodified single-machine applications over effectively unlimited remote computing resources. In this article, we present five serverless predictions for the next decade that will realize this vision of transparency -- equivalent to Tim Wagner's Serverless SuperComputer or AnyScale's Infinite Laptop proposals.

\end{abstract}
\maketitle

\chapterinitial{Serverless} gained popularity in industry and in academia in last few years \cite{definition}. It attracted companies and developers with a simple Function-as-a-Service (FaaS) programming model that realized original promise of the Cloud: elasticity and fine grained pay-as-you-go for actual usage. When a code is running in FaaS developers do not have control over where the code is running and do not need to worry about how to do scaling - serverless cloud providers create transparency by removing servers, or at least making them more transparent.

Transparency is an archetypal challenge in distributed systems that has not yet been adequately solved. Transparency implies the concealment from the user and the application programmer of the complexities of distributed systems.  According to Colouris \cite{colouris},  access  transparency  enables  local  and  remote  resources  to  be  accessed  using  identical operations. 

Nevertheless, Waldo et al. \cite{waldo} explain that the goal  of merging  the programming and computational models of local and remote computing is not new.  They state that around every ten years ``a furious bout of language and protocol design takes place and a new distributed computing paradigm is announced``. In every iteration, a new wave of software modernization is generated, and applications are ported to the  newest and hot paradigm. 

We believe the Serverless Compute paradigm, as emerging today \cite{definition,berkeley}, will converge at the needed level of resource abstraction to enable transparency. What we call {\it The Serverless End Game} is the process of mapping this principle on emerging disaggregated computing resources (compute, storage, memory), eventually enabling unlimited flexible scaling.

The major hypothesis of this paper is that full transparency will become possible in the next years thanks to predicted advances on latency reduction in distributed systems \cite{latency,attack,nvm}. This will put an end to the aforementioned cycles of software modernization. The consequences for the field will be enormous, by considerably simplifying development and maintenance of software systems for the majority of users.

\section{BACKGROUND}

Latency improvements \cite{latency,attack}  are boosting resource disaggregation in the Cloud, which is the definitive catalyst to achieve transparency. As we can see in table 1, current data center networks already enable disk storage disaggregation, where  reads from local disk are comparable (10ms) to reads over the network.  In contrast, creating a thread in Linux takes about 10{\textmu}s, still far better than 15ms/100ms (warm/cold) as achieved today in Function-as-a-Service (FaaS) settings. The level of resource disaggregation as possible today is specifically utilized by Serverless Platforms, and focus of research in Disaggregated Data Centers (DDC) in general \cite{disaggregation}.

\begin{table}
\caption{Latencies for remote resource access}
\label{table}
\small
\begin{tabular}{|p{28pt}|p{28pt}|p{51pt}|p{56pt}|}
\hline
Resource &  
Local & 
Remote today & 
Remote soon \\
\hline
Storage & 10ms & 10ms & 0.05ms (NVM)\\
Compute &  0.01ms (Thread) & 15-100ms (FaaS) & 0.001-0.1ms (RPC)\\
Memory & 0.0001ms &  0.25ms  (Redis) & 0.002-0.01ms (PMEM) \\
\hline
\end{tabular}
\label{tab1}
\end{table}


Providing access transparency over DDC resources is the aim of LegoOS: A disseminated, distributed {OS} for hardware resource disaggregation \cite{legoos}. 
LegoOS exposes a distributed set of virtual nodes (vNode) to users. Each vNode is like a  virtual  machine managing its own disaggregated processing, memory and storage resources.   LegoOS achieves transparency and backwards compatibility by  supporting  the  Linux  system  call  interface, so that  unmodified Linux applications can run on top of it.  For example, LegoOS executes two unmodified applications: Phoenix (a single-node multi-threaded implementation of MapReduce)  and TensorFlow. 

A good example of providing access transparency over serverless resources is Lithops \cite{lithops}. Lithops intercepts Python language libraries (multiprocessing) in order to access remote serverless resources in a transparent way. Lithops is however limited to running Python aplications using that library.



Another example of transparency in a serverless context is Faasm \cite{faasm}. Faasm exposes a specialized system interface which includes some POSIX syscalls, serverless-specific tasks, and frameworks such as OpenMP and MPI. Faasm transparently intercepts calls to this interface to automatically distribute unmodified applications, and execute existing HPC applications over serverless compute resources.

Faasm allows colocated functions to share pages of memory and synchronizes these pages across hosts to provide distributed state. However, this is done through a custom API where the user must have knowledge of the underlying system, hence breaking full transparency. Furthermore, when functions are widely distributed, this approach exhibits performance similar to traditional distributed shared memory (DSM), which has proven to be poor without hardware support.

Nevertheless, resource disaggregation is still in its infancy, and there is no current solution to provide flexible scaling and access transparency over remote shared memory. Container instantiation is slow comparing to local threads, and even fast NVMs\cite{nvm} are  an order of magnitude slower than local memory accesses which are  in the nanosecond range \cite{attack}.

Besides of the aforementioned constraints of current resource disaggregation, serverless 
computing has a number of well known limitations\cite{berkeley} like: focus on stateless computations, lack of efficient communication between executed tasks or functions, maximum code runtime limitations, and deficiencies in transparent integration of hardware accelerators.

\section{PREDICTIONS}


The major hypothesis of this paper is that transparency will be achieved in the next ten years thanks to novel advances in networking, disaggregation, and middleware services. The huge consequence is the unification of local and remote paradigms, which will democratize distributed programming for a majority of users. This will realize the old and ultimate goal of hiding the complexity of distributed systems.

The projected developments to reach the ultimate goal (Serverless End Game) include the following:
\begin{itemize}

\item Prediction 1: Serverless Clusters (Multi-tenant Kubernetes) will overcome the current limitations of  direct communication among functions, hardware acceleration, and  time limits.

\item Prediction 2: Serverless Granular computing will offer 1-10 {\textmu}s microsecond latencies  for remote functions thanks to lightweight virtualization and fast RPCs.

\item Prediction 3: Serverless memory disaggregation will  offer shared mutable state and coordination  at  2-10 {\textmu}s  microsecond latencies  over persistent memory. 

\item Prediction 4: Serverless Edge Computing platforms leveraging 6G's ms latencies  and AI optimizations will facilitate a Cloud Continuum for remote applications.

\item Prediction 5: Transparency will become the dominant software paradigm for most applications, when computing resources become standardized utilities.

\end{itemize}

We will discuss the basis of these predictions
as well as technical challenges
and risks. As we can see in Figure 1, the predictions are mapped to phases to reach the final goal. Let’s review the proposed forecasts.

\begin{figure}
\centerline{\includegraphics[width=18.5pc]{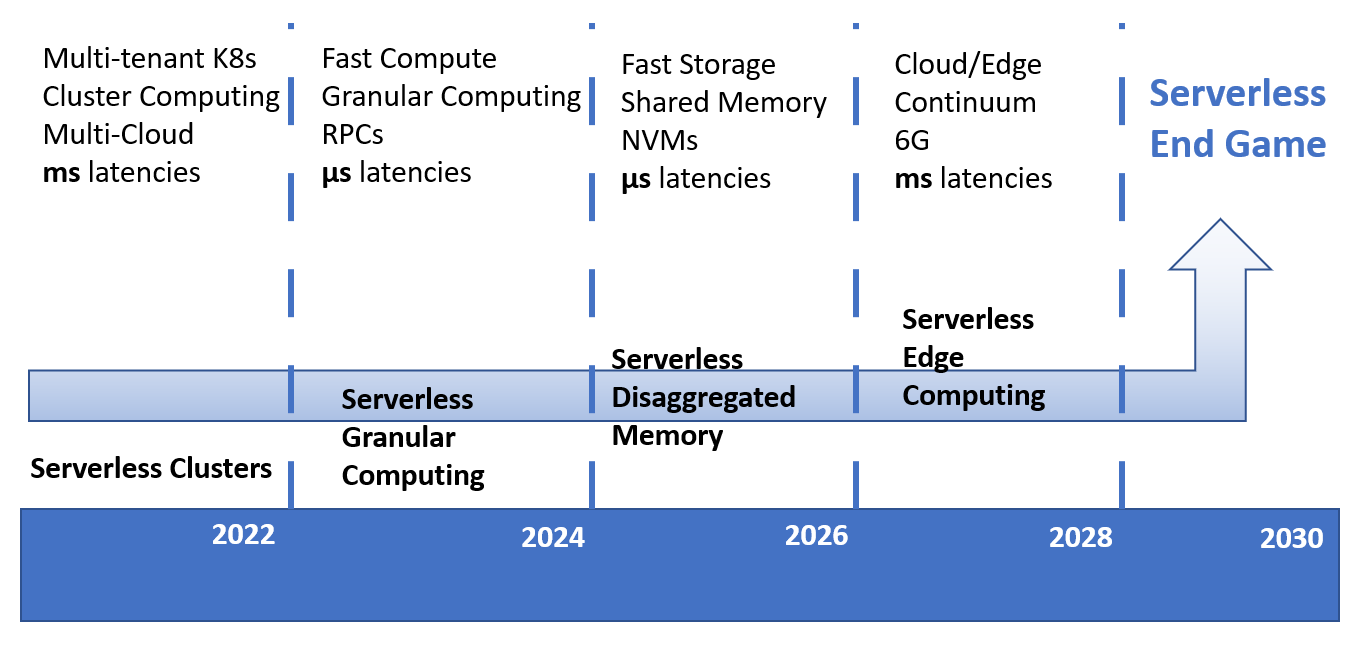}}
\caption{Serverless End Game}
\end{figure}

\subsection{Prediction 1}

In less than two years, a second generation of Serverless platforms will overcome major limitations of previous FaaS offerings. 

This is already starting to happen around the so called serverless containers platforms like Google Cloud Run, IBM Code Engine, or Amazon Fargate. Multi-tenant Kubernetes (K8s) clusters may become the necessary glue to build multi-Cloud API-agnostic applications. 

As a consequence, a number of Cluster toolkits will accelerate the transition to K8s like  Web frameworks (Flask, Rails, Node.js), Microservices (Quarcus, Spring, Micronaut), Analytics (Spark, Ray, Dask), Actors (Akka), and even High Performance Computing (MPI).

There will be intense efforts in cluster scheduling and workload analysis using machine learning techniques to optimize the deployment of different applications and frameworks. 

Nevertheless, container instantiation  times will still be slow (tens to 100s ms) and similar to current FaaS technologies. And the lack of memory disaggregation will preclude the transition for many applications.

\subsection{Prediction 2}

Serverless Granular Computing will drastically reduce  instantiation and execution times thanks to  more lightweight virtualization and execution  technologies\cite{granular} and fast RPCs\cite{rpc1}.  Microsecond latencies will also propagate to other middleware services such as messaging and collective communication.

Accessing disaggregated resources in a transparent manner requires a form of lightweight, flexible virtualization that does not currently exist. This new virtualization must intercept computation and memory management to provide access to disaggregated resources, and must do so with native-like performance and no input from the programmer. Several alternatives for lightweight virtualization will emerge around micro-VMs, software-based virtualization (Krustlet and WebAssembly) and platform-independent runtimes like GraalVM.

In summary, serverless granular computing will represent a performance boost for most applications in the Cloud, provoking a massive redesign of many toolkits (Web, games, Desktop). 

\subsection{Prediction 3}

 In the next years, public Cloud providers will start offering Serverless Memory Disaggregation  at microsecond speeds. We leverage here previous work on memory disaggregation like \cite{pocket, farm,infiniswap}.  Services like Serverless Redis will arrive soon with limited consistency requirements. 
 
Nevertheless, disaggregated memory at microsecond latencies will never reach the nanosecond latencies of local memory.  As a consequence, the combination of local memory with remote consistent shared memory will then create truly Serverless Mutable State and Coordination services. This will open transparency avenues for
  most stateful applications like: Serverless In-Memory Databases, Analytics (TensorFlow, PyTorch, Phoenix, OpenMP), Game engines and Multi-user Worlds (Unity, Minecraft), Desktop applications (Image editing, Video Editing), and  Collaborative applications (games, conferencing).

When Serverless Granular Computing and Disaggregated Memory become available, the three types of resources (compute, storage, memory) may be offered through virtual interfaces against disaggregated Serverless services. 

This will imply parallel work in different layers of the software stack. The first approach, as followed by Lithops\cite{lithops} will consist in intercepting language libraries and toolkits to transparently interact and combine local and remote resources.

The second approach will adapt polyglot execution runtimes like GraalVM, WASM, or .NET Common Language Runtime among others. This second phase is more ambitious, and it will consistently arrive to all languages and applications built on top of them.

Finally, the more advanced but also more complex approach is to redesign the Kernel Space to accomodate Serverless Remote Resources in a seamless way. This is more in line with initial DDC works like LegoOS\cite{legoos} or Arrakis\cite{arrakis}, but they will certainly require hard engineering efforts to cover the entire kernel APIs, and probably novel co-designs and OS improvements at all levels. We outline here the virtual file system and network device drivers, virtual memory management, and also process/thread management layers.

In this line, Arrakis\cite{arrakis} comes from previous efforts aimed at optimizing the kernel code paths  to improve data transfer and latency in the OS. In Arrakis, applications have direct access to virtualized I/O devices, which allows most I/O operations to bypass the kernel entirely without compromising process isolation. Arrakis virtualized  control plane approach  allows storage solutions to be integrated with applications (co-design), even allowing the development of higher level abstractions like persistent data structures.  Arrakis shows us how its control plane may become a first step towards integration with a serverless data center resource allocator.

Redesign of OS kernels will then enable the final goal of full transparency for any native applications using OS public Call Interfaces. This will require the collaboration of DDC and OS researchers, but also the research work of the broader Systems community.

\subsection{Prediction 4}

The computing power at the edge, along 1-2 millisecond latencies facilitated by 6G will produce Serverless Edge platforms that will alter the operating systems of mobile phones and terminals at the edge. The smooth integration with Cloud resources will create a  Cloud Continuum where transparent applications directly access data in remote resources.

Micro-second latencies will also produce a flourishing of remote use interfaces. In line with remote displays (X-Windows, VNC), many applications could just focus edge devices in user interface interaction, moving all computing logic to remote execution.
This could provoke a revival of dumb terminals that delegate their execution entirely to the cloud.

Nevertheless, dumb terminals cannot address more sophisticated use cases where edge computing resources are relevant. In cases where you cannot neglect the computing resources in the edge, local and remote resources should be intelligently combined in the Edge/Cloud Continuum.

This will certainly become the last frontier for software modernization, which will then define how to write local  applications that use remote resources in a transparent way. A plethora of novel applications (tele-presence, games, collaborative work, augmented/mixed reality, virtual reality) will emerge leveraging the novel programming abstractions, and existing ones will be re-engineered or improved considerably.

\subsection{Prediction 5}

In ten years, no application domain will remain unaltered for transparency as a consequence of the Serverless End Game. The convergence of  Granular Computing, Disaggregated memory, and the Edge-Cloud continuum will definitely unify the local and remote paradigms. Computing will finally become a utility thanks to standards accepted by all providers. 


\section {IMPACT}

Achieving transparency will produce a profound impact in the computing field and it will change drastically how applications are built in the future.

On the technical side, transparency has been the ultimate goal of distributed systems for many decades now. Developers will still care about parallel programming, object oriented programming, user interfaces, event-based systems, or functional programming. However, most developers will no longer worry about middleware, transport layers and marshaling, web-based protocols, or Cloud-specific APIs among others.

On the technical side, the disaggregation (and standardization) of remote computing resources will also have a profound impact on the edge: in the redesign of Operating Systems for computers, mobile phones, and edge devices. Novel development environments will emerge to cover the entire life cycle of applications in the Cloud/Edge continuum.

On the economic side, transparency will also imply a revolution that will shake the entire industry. The major economic impact will be in developer costs and productivity for most applications. If Serverless technologies transparently handle the distributed systems complexity, programming remote applications will become as simple as programming local ones with infinite resources. 

Transparency will boost the consumerization of software, facilitating the creation of many applications in different domains. More users will be able to customize and develop software,  benefiting from novel visual and no-code tools.

The economic impact will also affect all Cloud providers on a first wave, in a fierce competition to be the most efficient disaggregated backend for applications. It will also affect Operating Systems in the edge (like Apple, Windows, Linux, Android, or iOS), which will  seamlessly integrate remote resources.

In this integration of Cloud and Edge, some providers with presence in both sides (Google, Microsoft, Apple) will have a certain advantage over the rest of the industry. Nevertheless, open source giants like IBM/RedHat could also offer integrated solutions in the Cloud and Edge based on Linux technologies.

In any case, microsecond latencies enabled in the data center, and 1-5 ms latencies of edge resources will profoundly change how we interact with technology. Analogously to the unification of the local and remote paradigms in programming, we will see a blurring and unification of real and virtual environments using Augmented and Virtual Reality.


\section{TRADEOFFS OF DISAGGREGATION AND TRANSPARENCY}

Blurring the borders between local and remote resources may have important advantages, but it also implies some tradeoffs and limitations that we must be aware. 

Waldo et al.\cite{waldo} already considered 25 years ago that future hardware improvements could make the difference in latency irrelevant,  and that differences between local and remote memory could be masked.  But they still claimed that concurrency and partial failures preclude the unification of local and remote computing. Let´s study the major challenges and tradeoffs:

\begin{enumerate}

\item Partial Failure:  Existing applications are a  blackbox for the Cloud, but the transition will imply a ``compile to the Cloud`` process. In this case, the Cloud will have access  over applications' life cycle and it will be able to offer failure transparency, and to optimize their execution performance and cost.  This means that they can  perform static analysis to predict resource requirements, failures, dependencies and potential for hardware acceleration.  In the past, extensive optimizations have been performed to declarative dataflows, but there are also opportunities to improve imperative programs with static analysis \cite{kiran2009execution}.


Transparency efforts for different types of applications will also require customizable control planes. In particular, customized co-designs \cite{angel} may be necessary to  optimize resource usage. Such customization will be based on  advanced observability and fast orchestration mechanisms relying on standard services and protocols.  Monitoring and interception of the different resources (compute, storage, memory, network) should be available and even integrated into the data center, enabling coordinated actuators at different levels.

\item Concurrency:  Another important limitation when developing concurrent applications is scaling transparency, which means that applications can expand in scale without changes to the system structure or the application algorithms. If the local programming model was designed to use a fixed amount of resources, there is no magic way of transparently achieving scalability, not to mention elasticity. 

Some workloads that do not need elasticity, such as enterprise batch jobs or scientific simulations, can just port existing code to the Cloud. If the application is configured to run on a fixed number of resources (threads or processes), it may run unmodified over the same fixed amount of disaggregated resources.  However, for more user driven and interactive services, such as internal enterprise web applications, simple porting of the executables (sometimes referred as ``lift-and-shift``) is rarely enough. The unchanged code is not able to take advantage of the elasticity of disaggregated resources.

We will then need elastic programming models that can be used without change when running over Cloud resources. Such elastic models should take care of providing the different  transparency types (scaling, failure, replication, location, access) and other aspects of application behavior when it is moved between local and distributed environments.  The local executable APIs may need to be expanded to include elastic programming abstractions for processes, memory, and storage.


\item Locality:  Advances in datacenter networking and NVMs have reduced access to networked storage to a few 10ths of  {\textmu}s, however this is still an order of magnitude slower than local memory accesses which are  in the nanosecond range \cite{attack} (100ns), and local cache accesses in the 4ns-30ns range. Existing efforts in memory disaggregation  \cite{farm, infiniswap} strive to play in the {\textmu}s range, which can be a limiting factor for some applications. This means that local memory cannot be neglected, since it will always offer an order of magnitude faster access than remote memory. Future approaches should smartly leverage local memory and combine it with remote memory.

Locality still plays a key role in stateful distributed applications.  For example:  (i)  where huge data movements represent a penalty and memory-locality can be useful; (ii) where specialized hardware like GPUs must be used; in (iii) some iterative machine-learning algorithms; in (iv) simulators, interactive agents or actors.

Future serverless middleware addressing transparency will have to provide affinity and grouping requirements for stateful entities. Serverless Stateful services will support very different requirements of coordination, consistency, scalability and fault tolerance.

\item Cost: Another limiting factor today for current serverless disaggregated  technologies is cost. For example, Amazon Lambda compute time is 2x more expensive than on-demand VMs, and around 6x more expensive than Spot Instances. For problems like batch analytics that have 100\% utilization, VMs are now a better solution for intensive computing applications. 

We foresee that cost limitations will be overcome in the future, when Cloud providers improve scheduling of lightweight micro-VMs, reducing both start time and overall costs thanks to locality. Furthermore, sophisticated cost planes should be available to users in order to have full control of costs derived by remote resources.

\end{enumerate}

\section{CONCLUSIONS}

We argue that full transparency will be possible soon thanks to low latency and resource disaggregation. The Serverless End Game will unify local and remote programming paradigms, changing completely the way we currently create distributed applications. This is the ultimate goal of distributed systems, to become invisible using transparent middleware, and to simplify how users access remote resources.

\section{ACKNOWLEDGMENT}

This work has been partially supported by the EU Horizon2020 programme under grant agreement No 825184.

\begin{IEEEbiography}{Pedro Garcia Lopez}{\,}is full professor at  Universitat Rovira i Virgili (Spain) and visiting scientist at IBM Watson Research. He leads the “Cloud and Distributed Systems Lab” research group and coordinates the european research project "CloudButton: Serverless Data Analytics". Contact him at pedro.garcia@urv.cat.
\end{IEEEbiography}

\begin{IEEEbiography}{Aleksander Slominski}{\,}  is research staff member in the Serverless Group in Cloud Platform, Cognitive Systems and Services Department at IBM T.J. Watson Research Center in Yorktown Heights, NY, USA. Contact him at \url{https://aslom.net}
\end{IEEEbiography}

\begin{IEEEbiography}{Michael Behrendt}{\,} is a Distinguished Engineer and  technical executive at IBM Deutschland Research. He leads key Serverless services like IBM Cloud Functions and IBM Code Engine. Contact him at michaelbehrendt@de.ibm.com.
\end{IEEEbiography}

\begin{IEEEbiography}{Bernard Metzler}{\,}is a Principal Research Staff Member and Technical Leader at IBM Zurich Research Laboratory. His main research interests
are in enhancing network and storage IO of distributed systems,
and the integration of modern high performance IO hardware with distributed
applications.  Contact him at bmt@zurich.ibm.com.
\end{IEEEbiography}

\end{document}